\documentclass[aps,pra,reprint,groupedaddress]{revtex4-1}
\usepackage{graphicx}
\usepackage{amsmath}
\usepackage{amssymb}
\usepackage{bm}
\usepackage{braket}
\usepackage{graphicx}
\usepackage{comment}
\usepackage{xr}
\usepackage{xcolor,soul}
\usepackage{siunitx}
\usepackage[UKenglish]{babel}
\usepackage{bm} 
\renewcommand{\Re}{\operatorname{Re}}
\renewcommand{\Im}{\operatorname{Im}}

\begin{document}
\renewcommand{\thefootnote}{\fnsymbol{footnote}}

\title{Low-loss Plasmonic Dielectric Nanoresonators}


\author{Yi Yang$^1$}
\email{yiy@mit.edu}
\author{Owen D. Miller$^2$}
\email{owen.miller@yale.edu}
\author{Thomas Christensen$^1$}
\author{John D. Joannopoulos$^1$}
\author{Marin Solja\v{c}i\'{c}$^1$}
\affiliation{$^1$Research Laboratory of Electronics, Massachusetts Institute of Technology, Cambridge, MA 02139, USA \\$^2$Department of Applied Physics, Yale University, New Haven, CT 06520, USA}

\begin{abstract}
    Material losses in metals are a central bottleneck in plasmonics for many applications. Here we propose and theoretically demonstrate that metal losses can be successfully mitigated with dielectric particles on metallic films, giving rise to hybrid dielectric--metal resonances. In the far field, they yield strong and efficient scattering, beyond even the theoretical limits of all-metal and all-dielectric structures. In the near field, they offer high-Purcell-factor (${>}5000$), high-quantum-efficiency (${>}90\%$), and highly directional emission at visible and infrared wavelengths. Their quality factors can be readily tailored from plasmonic-like (${\sim}10$) to dielectric-like (${\sim}10^3$), with wide control over the individual resonant coupling to photon, plasmon, and dissipative channels. Compared with conventional plasmonic nanostructures, such resonances show robustness against detrimental nonlocal effects and provide higher field enhancement at extreme nanoscopic sizes and spacings. These hybrid resonances equip plasmonics with high efficiency, which has been the predominant goal since the field's inception.
\end{abstract}

\maketitle
\sisetup{range-phrase=--}
\sisetup{range-units=single}

The material composition of an optical nanoresonator dictates sharply contrasting properties: metallic nanoparticles~\cite{novotny2011antennas,giannini2011plasmonic,biagioni2012nanoantennas,tsakmakidis2016large} support highly subwavelength plasmons with large field strengths, but which suffer from intrinsic material losses~\cite{PhysRevLett.97.206806,PhysRevLett.110.183901,PhysRevLett.112.123903,miller2016fundamental,miller2015shape}, whereas dielectric nanoparticles~
\cite{krasnok2012all,kuznetsov_optically_2016,jahani2016all} support exquisite low-loss versatility, but only moderate confinement as their sizes must generally be wavelength-scale or larger. In this Article, we propose and theoretically demonstrate that a combined approach---dielectric nanoparticles on metallic films---can exhibit a unique combination of strong fields and high confinement alongside small dissipative losses. We show the utility of such hybrid plasmonic dielectric resonators for (i)~far-field excitations, where subwavelength silicon-on-silver nanoparticles can scatter more efficiently than is even theoretically possible for any all-metal or all-dielectric approach, and (ii)~near-field excitations, where highly directional spontaneous emission enhancements ${>}5000$ are possible with quantum efficiencies ${>}90\%$ and even approaching unity. Moreover, the dielectric composition of the nanoparticle, when placed atop a metallic supporting film, should mitigate much of the quantum- and surface-induced nonlocal damping that occurs at nanometer scales, an effect we confirm quantitatively with a hydrodynamic susceptibility model. Furthermore, as our approach does not rely on nanostructured metallic components, it strongly constrains parasitic dissipation arising from fabrication imperfections.
More broadly, simple geometrical variations provide wide control over the individual resonant-coupling rates to photon, plasmon, and dissipative degrees of freedom, opening a pathway to low-loss, high-efficiency plasmonics.

Mitigating loss is a pivotal goal~\cite{khurgin2015deal,tassin2012comparison,boltasseva2011low,naik2013alternative} in plasmonics. When nanoparticles interact with plane waves, their cross-sections are typically dominated by dissipative absorption. In the near field, large spontaneous-emission enhancements (Purcell factors) have been demonstrated~\cite{rogobete2007design,kinkhabwala2009large,akselrod2014probing,eggleston2015optical} through mode-volume squeezing, but it has been typically accompanied by sub-50\% quantum efficiencies at visible frequencies. Hybrid structures~\cite{devilez2010compact,rusak2014hybrid} with separated dielectric (director) and metal (feed) functionality have been proposed for better radiative efficiency, but with lower enhancements. This tradeoff gives the appearance that strong and efficient plasmonic antennas are only possible at infrared frequencies~\cite{khurgin2015deal}, where they behave akin to perfect conductors and ``plasmonic'' effects are minor. Quantum corrections in plasmonics~\cite{feibelman1982surface,zhu2016quantum,christensen2016quantum}, e.g. due to electron tunneling~\cite{savage2012revealing} and nonlocality~\cite{garcia2008nonlocal}, further limit the ultimate enhancement of plasmonic resonators.

The difficulty of achieving low-loss plasmons has led to the perception that high confinement is simply incompatible with low loss, as large fields near/in a metal surface may necessarily generate significant dissipation. This intuition has led to the burgeoning field of alternative plasmonic materials~\cite{boltasseva2011low,naik2013alternative}, whereby highly doped semiconductors or polar dielectrics ideally exhibit negative real permittivities with small imaginary (lossy) parts. There has been a complementary effort in all-dielectric nanoparticles~\cite{krasnok2012all,kuznetsov_optically_2016} and metamaterials~\cite{kuznetsov_optically_2016,jahani2016all}, but subwavelength resonances fundamentally require metallic components with negative permittivities~\cite{khurgin2015deal,PhysRevLett.97.206806}. Material engineering has also been proposed in the form of band engineering~\cite{khurgin2010search} and gain offsets~\cite{zayats2013active}. The perceived confinement--loss tradeoff is rigorously correct for quasistatic plasmonic resonators~\cite{PhysRevLett.97.206806}, in which the desired resonant frequency directly sets the fraction of the field intensity that must reside within the lossy metal~\cite{PhysRevLett.97.206806,PhysRevLett.110.183901,khurgin2012reflecting}. In closed non-radiative plasmonic systems, proper geometrical optimization of dielectric-metal waveguides can reduce propagation losses \cite{oulton2008hybrid};
%
%
in open systems, the central unanswered question is whether their radiative coupling rates can be strongly increased such that radiation significantly exceeds near-field dissipative losses. Here we show that open resonators comprising high-index, low-loss nanoparticles on metallic films can simultaneously achieve high confinement and high radiative efficiencies, without significant dissipative loss.



\begin{figure}[htbp]
	\centering
	\includegraphics[width=0.85\linewidth]{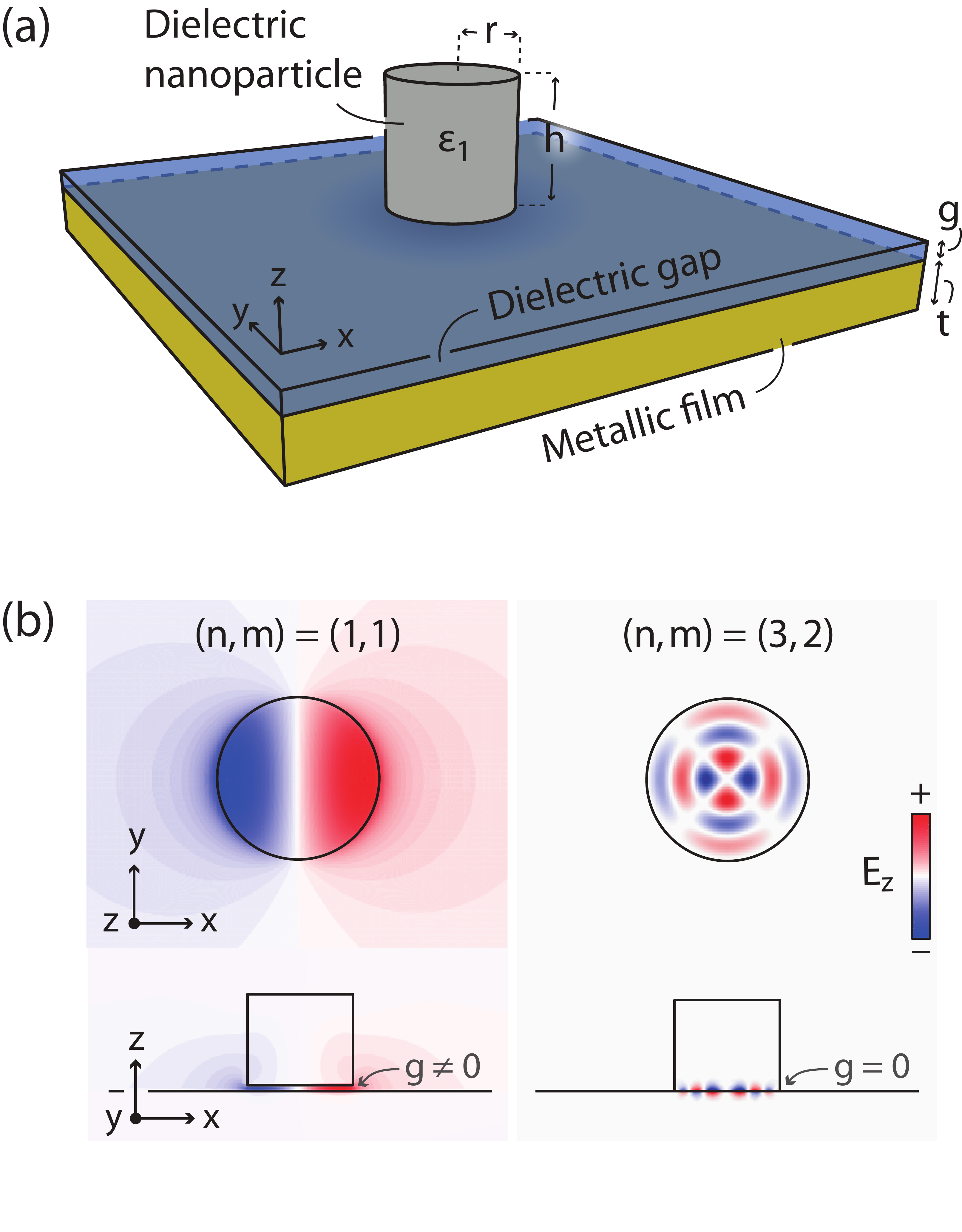}
        \caption{\textbf{Hybrid dielectric--metal resonances.} (a) Schematic of the structure, composed of a metallic layer of thickness $t$, a dielectric spacer with gap size $g$, and a dielectric cylindrical nanoparticle of permittivity $\varepsilon_1$, height $h$, and radius $r$. For simplicity, we here consider vacuum as the ambient and gap media. (b) $E_z$ mode profiles of two selected hybrid resonances, for a Si cylinder on a Ag substrate. (Material parameters detailed in Supplementary~I).
        }
	\label{schematic}
\end{figure}

\section{Conceptual Basis}

We propose a hybrid dielectric--metal resonator [Fig.~\ref{schematic}(a)] that mitigates restrictions from metal losses on plasmonic scattering, emission, and quality factors to a great extent. The cylindrical symmetry implies that resonances can be labeled with indices $(n,m)$, enumerating field variations in the radial and azimuthal directions, respectively. Unlike the widely used all-metal ``gap-plasmon'' resonances~\cite{tsakmakidis2016large,akselrod2014probing} (hereafter, metal--metal resonances), which require a nonzero gap to squeeze the field inside, the dielectric--metal resonances strongly confine the resonant field for either zero or nonzero gap [Fig.~\ref{schematic}(b)].
\begin{figure*}[htbp]
  \centering
  \includegraphics[width=.95\textwidth]{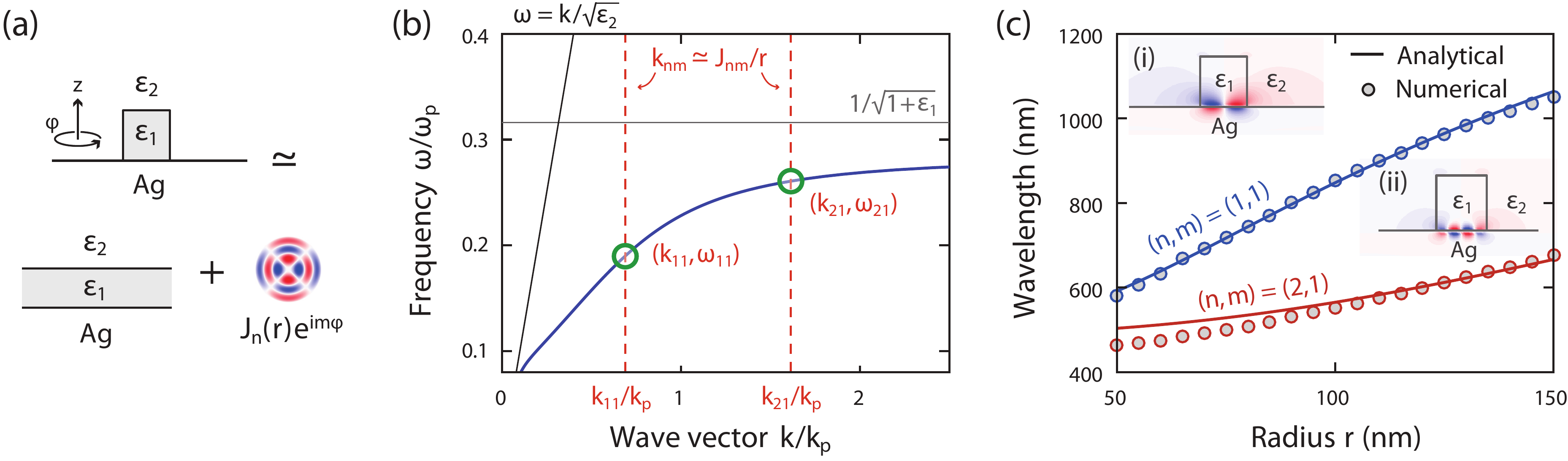}
    \caption{\textbf{Analytical model of hybrid dielectric--metal resonances.} (a)~Pictorial representation of the hybrid resonance, which approximately satisfies a Bessel-function phase-matching condition, Eq.~\ref{ana}, imposed on the underlying planar structure. (b)~The application of Eq.~\ref{ana} illustrated in a concrete system ($h = \SI{100}{nm}$, with $\varepsilon_1$ = 12.25, $\varepsilon_2$ = 1, $t=\infty$, and $g=0$): the underlying planar system's plasmon dispersion (blue) and the resonant wavevectors $k_{nm}$ (red dashed) dictate resonant frequencies $\omega_{nm}$. $\omega_{\rm p}$ and $k_{\rm p}$ denote plasma frequency and $k_{\rm p}=\omega_{\rm p}/c$ ($c$ being speed of light). (c)~The resonant wavelengths of the $(1,1)$ and $(2,1)$ [$E_z$ profiles shown in (i) and (ii) respectively] modes versus cylinder radius $r$, as predicted by Eq.~\eqref{ana} (solid lines) and numerical computations (circles).
}
  \label{analytics}
\end{figure*}

Conceptually, the dielectric--metal resonances can be understood as the surface plasmons of a planar multilayer metal--dielectric system restricted to specific quantized wavevectors $k_{nm}$. The nanoparticle's boundary reflects surface plasmons of general wavevector $k$ without phase shift. For a cylinder of radius $r$, the round-trip phase over the nanoparticle is given by the Bessel function of the first kind $J_{n}(kr)$. Localized resonances are supported when this round trip phase vanishes, i.e., at the Bessel zeros $J_{nm}$:
\begin{equation}
k_{nm}r \simeq J_{nm}. \label{ana}
\end{equation}
Resonant frequencies are obtained by sampling the multilayer surface plasmon dispersion curve, $\omega(k)$, at the resonant wavevectors $k_{nm}\simeq J_{nm}/r$ [Fig.~\ref{analytics}(b)], as verified by the agreement between analytics and numerics [Fig.~\ref{analytics}(c)]. Eq.~\eqref{ana} is most accurate for low-order resonances, when the plasmon reflection phase~\cite{gordon2006light} at the nanoparticle boundary is small ($\Re{k}\gg \Im{k}$). Eq.~\eqref{ana} is also generalizable to other nanoparticle geometries and more complex multilayers.

This simple, yet accurate picture of the hybrid resonances, as part-plasmon, part-Bessel resonances, illustrates the separation of key functionality: the plasmonic metal provides vertical confinement, while the dielectric provides horizontal confinement and dictates the resonant condition. External radiative coupling occurs at the low-loss dielectric--air interface, away from the lossy metal, enabling higher radiative efficiencies than those in conventional plasmonic nanostructures.

\section{Far-Field Scattering}

Metallic nanoparticles generally scatter more strongly than all-dielectric nanoparticles. Yet this large scattering strength---as measured, e.g., by the optical cross-section per unit particle volume---is typically accompanied by significant absorption. Thus for many applications where absorption is undesirable (such as photovoltaics~\cite{atwater2010plasmonics}), the critical figure of merit is scattering strength accompanied by high radiative efficiency. Here we leverage recently developed optical-response bounds to show that low-loss dielectric nanoparticles on metallic films can achieve subwavelength scattering with large radiative efficiency, surpassing all-metal and all-dielectric scatterers and approaching fundamental limits.

There has been significant interest in finding general upper bounds to optical response~\cite{gustafsson2007physical,hugonin2015fundamental}, and recently we developed new such bounds~\cite{PhysRevLett.112.123903,miller2016fundamental,miller2015shape}. Passivity, which requires non-negative absorbed and scattered powers, imposes limits to the currents that can be excited in an absorptive scatterer, leading to bounds that are independent of shape, which account for material loss ($\propto \Im\chi$, for material susceptibility $\chi$), and which can incorporate radiative-efficiency constraints. The bounds demonstrate~\cite{miller2016fundamental} that high radiative efficiency, defined as $\eta \equiv \sigma_{\rm sca}/ \left(\sigma_{\rm sca} + \sigma_{\rm abs}\right)=\sigma_{\rm sca}/ \sigma_{\rm ext}$ (where $\sigma_{\rm sca}$, $\sigma_{\rm abs}$, and $\sigma_{\rm ext}$ are the scattering, absorption, and extinction cross sections, respectively), necessarily reduces the largest cross-section per volume that can be achieved. A natural figure of merit (${\rm FOM_{sca}}$) emerges: $\sigma_{\rm sca}/V\times 1/[\eta(1-\eta)]$ (equivalently, $\sigma_{\rm ext}/\sigma_{\rm abs}\times\sigma_{\rm ext}/V$), which rewards high scattering cross-section ($\sigma_{\rm sca}/V$) as well as high radiative efficiency ($\eta \gg 0.5$). The $\rm FOM_{sca}$ is subject to the bound~\cite{miller2016fundamental}
\begin{eqnarray}
    \rm FOM_{\rm sca}\equiv\frac{\sigma_{\text{sca}}/V}{\eta \left(1 - \eta\right)} \leq \frac{\omega}{c}\frac{|\chi(\omega)|^2}{\Im\chi(\omega)} \frac{I_{\rm inc}}{I_0},
    \label{scaBound}
\end{eqnarray}
which depends only on the frequency $\omega$, the material composition, and the incident field properties. $I_{\rm inc}/I_0$ is the ratio of the incident-field intensity $I_{\rm inc}$ (including e.g., reflection from a planar film in the absence of the nanoparticle) integrated over particle volume to the intensity of the plane wave. Perfect radiative efficiency ($\eta=1$) is unachievable for lossy scatterers, such that Eq.~\eqref{scaBound} cannot diverge. Equation~\eqref{scaBound} clearly shows that low-loss materials offer the possibility for strong and high-efficiency scattering, but all-dielectric structures cannot reach their bounds (in most parameter regimes) for lack of subwavelength resonances. On the other hand, by equipping dielectric nanoparticles with a subwavelength resonant mechanism, achieved by coupling to a metallic substrate, these high limits may actually be approached.
\begin{figure}[hbtp]
  \centering
  \includegraphics[width=\linewidth]{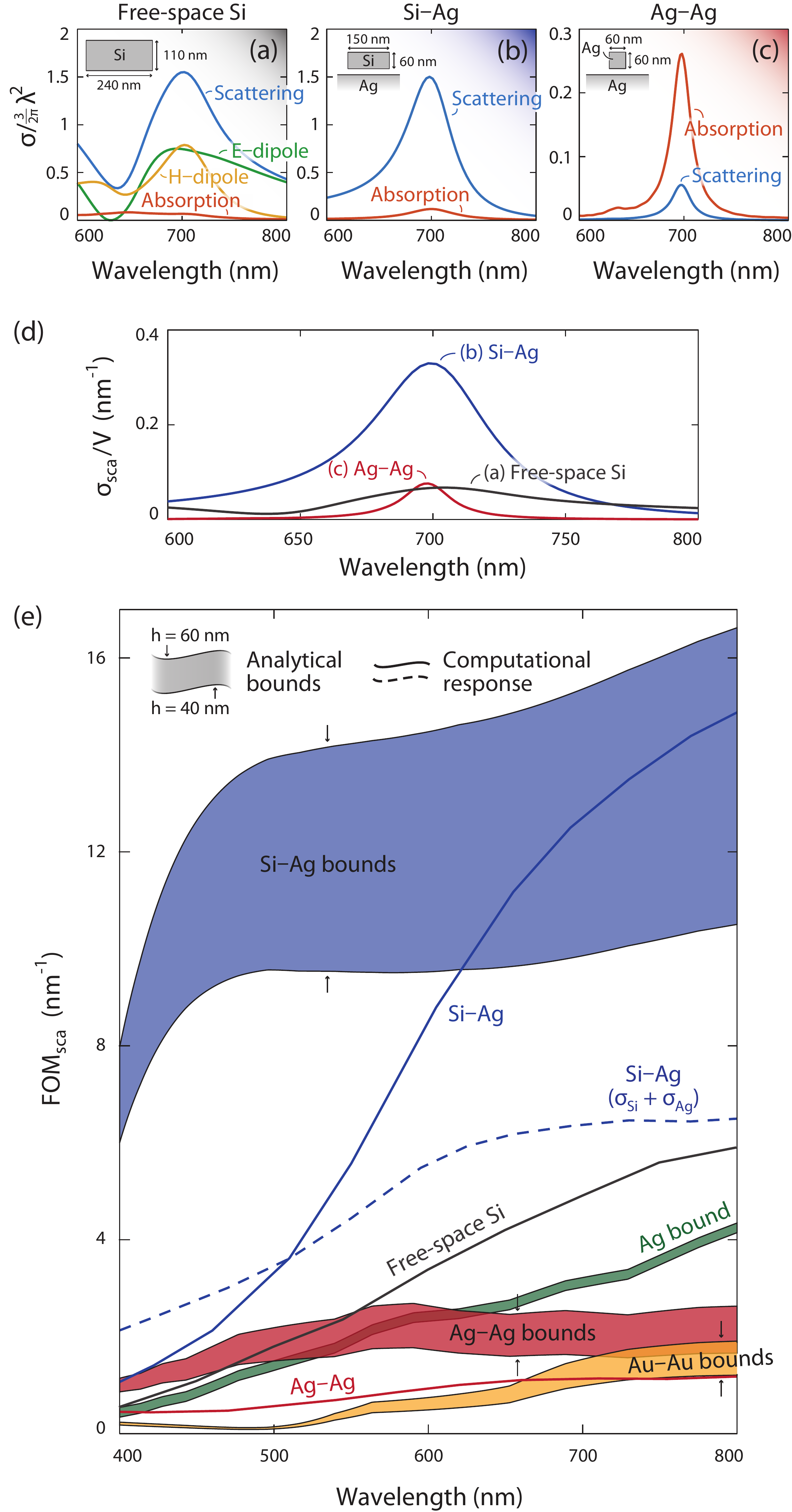}
    \caption{\textbf{Dielectric--metal resonances offer strong scattering accompanied by modest absorption, at combined rates that cannot be achieved by all-metal or all-dielectric structures.} Top: Scattering and absorption cross sections of nanoparticles under varying material and environment composition: (a) Si cylinder in free-space; (b-c) Si and Ag cylinders, respectively, above a semi-infinite Ag substrate with gap thickness $g=$\SI{2}{\nm}. Geometrical parameters (insets) are chosen to align their resonant wavelengths at \SI{700}{\nm}. The three structures are all illuminated by normally-incident plane waves. In (b-c), the absorption includes the dissipation in both the particle and the substrate.
    (d)~The dielectric--metal structure shows the highest per-volume scattering cross-section, because it simultaneously achieves large scattering cross-section $\sigma_{\rm sca}$, high radiative efficiency $\eta$, and a small particle volume $V$.
    (e)~In the visible regime, the scattering capabilities of metal--metal geometries (Ag--Ag and Au--Au bounds), free-space metallic (Ag bound), and free-space dielectric (Si free-space) scatterers all fall short when compared with the dielectric--metal (Si--Ag) scatterer, which also approaches its own upper bound, per Eq.~\eqref{scaBound}. For the Si--Ag and Ag--Ag structures, the gap size is fixed at \SI{5}{\nm}; the cylinder (both Si and Ag) height $h$ ranges from \SIrange{40}{60}{\nm} in order to tune the resonant wavelength.
}
  \label{scattering}
\end{figure}

We compare scattering by three types of resonators---(i) a free-space, all-dielectric resonator, (ii) a hybrid dielectric-on-metal resonator, and (iii) a metal-on-metal resonator---at \SI{700}{\nm} wavelength. For each resonator, the dielectric is Si. The free-space dielectric resonator [Fig.~\ref{scattering}(a)] is designed to achieve super-scattering~\cite{PhysRevLett.105.013901} (Supplementary~II), with $\eta\approx96\%$, via aligned electric- and magnetic-dipole moments. The hybrid silicon-on-silver resonator [Fig.~\ref{scattering}(b)] is optimized to have a similar scattering cross-section, which is achieved in roughly one-fifth of the volume and with $\eta\approx93\%$.  Finally, the radius of the Ag-on-Ag resonator [Fig.~\ref{scattering}(c)] is optimized by radius [cylinder height and gap size same as Fig.~\ref{scattering}(b) for constant $I_{\rm inc}$]; notably, it only achieves only $\approx17\%$ radiative efficiency. Figure~\ref{scattering}(d) compares the scattering strengths of the three architectures, measured by $\sigma_{\rm sca}/V$, clearly showing the dielectric--metal structure's advantage, which remains compelling across visible frequencies [Fig.~\ref{scattering}(e)]. Fig.~\ref{scattering}(e) compares $\rm FOM_{\rm sca}$ of different structures and includes corresponding bounds (shaded regions) based on the cylinder height (Supplementary~III) due to the oscillatory incident fields in the presence of the reflective film. Different from Fig.~\ref{scattering}(a-d), all cross-sections in Fig.~\ref{scattering}(e) (except the dashed line) isolate the contributions of the nanoparticles themselves (Supplementary~IV), and the substrate is incorporated in the incident-field definition~\cite{miller2015shape}. At longer wavelengths, the scattering strength of the Si cylinder (blue solid line) approaches its bound, the highest among all bounds. By replacing the cylinder with a horizontally-aligned nanorod in the dielectric--metal system, scattering bounds can be saturated across the entire visible spectrum (Supplementary Fig.~S1). Including film absorption and scattering in the dielectric-metal structure (blue dashed line), the hybrid resonance retains large $\rm FOM_{\rm sca}$, still outperforming all-metal and all-dielectric resonators. In the following section, we translate this large-response, high-radiative-efficiency capability from the far field to the near field.

\section{Near-Field Emission Enhancements}
\begin{figure*}[hbtp]
  \centering
  \includegraphics[width=0.6\linewidth]{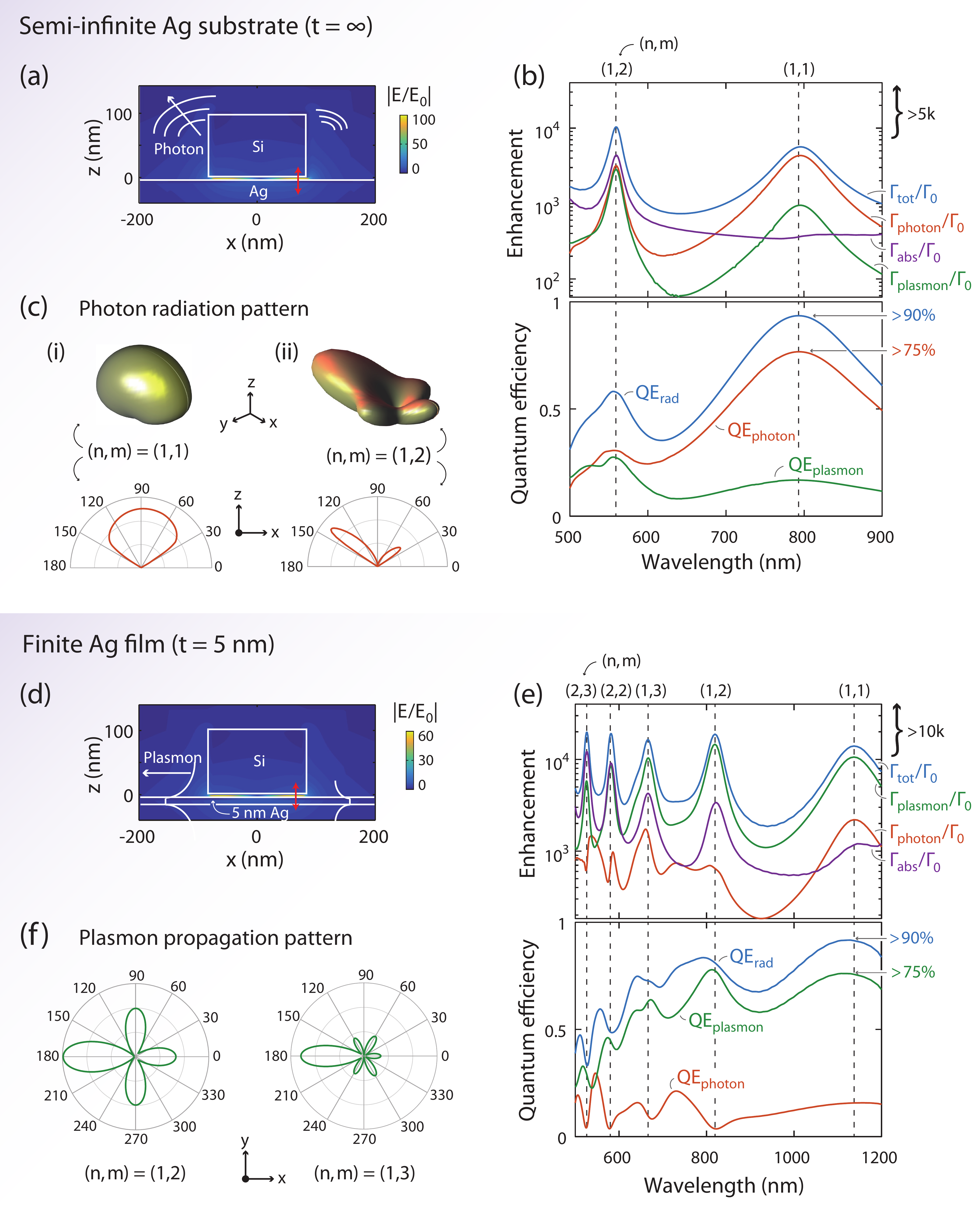}
  \caption{\textbf{High-Purcell, high-efficiency, high-directionality spontaneous emission enhancement with the hybrid resonances.}
    (a)~Structure and its $(1,1)$ modal profile for photon emission. An $r=\SI{80}{\nm}$, $h=\SI{100}{\nm}$ silicon cylinder above semi-infinite Ag with a $g = 2$~nm gap. A $z$-oriented dipole (red arrow) is located in the middle of the gap and at $x = \SI{67}{\nm}$.
 (b)~Enhancement decomposition reveals strong and efficient photon emission. A high quantum efficiency ${>}90$\% and photon efficiency ${>}75$\% are achieved using the (1,1) mode.
 (c)~Far-field photon radiation pattern of the $(1,1)$ and $(1,2)$ mode. Highly directional photon emission is achieved using the $(1,2)$ mode.
 (d)~Structure and its $(1,1)$ resonance profile for plasmon emission. A finite-thickness ($t=5\text{ nm}$) metallic film is considered; all other parameters mirror those in (a).
 (e)~Enhancement decomposition reveals strong and efficient plasmon launching. The $(1,1)$ mode achieves a total radiative efficiency ${>}90\%$ and a plasmon efficiency ${>}75$\%.
 (f)~Directional plasmon propagation with the $(1,2)$ and $(1,3)$ mode.
}
  \label{emission}
\end{figure*}

Plasmonic losses are particularly acute in the near field, for sources in close proximity to the resonator, as the source readily accesses lossy channels that dissipate energy before it can escape into a propagating far-field photon or guided plasmon. In contrast, with negligible local dissipation, dielectric--metal resonances can provide high-Purcell, high-efficiency, and high-directionality spontaneous emission enhancements. A Purcell factor ${>}5000$ with quantum efficiency (including both photon and plasmon emission) ${>}90\%$ can be achieved in the optical regime. Whereas some previous work (e.g., Ref.~\cite{akselrod2014probing}) has not distinguished between emission into guided plasmons and emission into radiating photons, we separate each contribution and show that a simple geometrical reconfiguration (increasing/reducing the metal-film thickness) can swing the emission rate from plasmon-dominant ($>75\%$) to photon-dominant ($>75\%$) or vice versa. Directional photon and plasmon emission can also be realized via high-order resonances.

We first demonstrate \emph{photon} emission enhancement with a silicon cylinder on a semi-infinite Ag substrate, separated by a \SI{2}{\nm} gap [Fig.~\ref{emission}(a)]. Planar dispersion analysis (Supplementary Fig.~S2) suggests that this geometry should provide similar Purcell enhancement, and much higher quantum efficiency, as compared to a \SI{5}{\nm}-gap-size metal--metal structure. We decompose~\cite{yang2016near} the enhanced emission from a $z$-oriented dipole into far-field photon, guided plasmon, and local dissipative channels and obtain corresponding efficiencies (Supplementary~VII) [Fig.~\ref{emission}(b)]. The $(1,1)$ and $(1,2)$ modes achieve Purcell factors (total enhancement) ${>}5000$ and ${>}10^4$, respectively. As importantly, the $(1,1)$ mode exhibits ${>}90\%$ quantum efficiency and ${>}75\%$ photon efficiency. Similar efficiencies are achieved for emitters located throughout the gap region (not shown; adopting the approach in~\cite{yang2016optically}). In the far field [Fig.~\ref{emission}(c)], the $(1,1)$ mode exhibits wide-angle emission, while the $(1,2)$ mode enables highly directional photon emission, without the Yagi-Uda configuration~\cite{devilez2010compact,curto2010unidirectional} or a periodic lattice~\cite{lozano2013plasmonics}.

Even higher quantum efficiencies,  with similar enhancements, are possible with alternative low-loss dielectric materials (on Ag). AlSb~\cite{zollner1989dielectric} nanoparticles offer close-to-unity efficiencies below their \SI{2.2}{\eV} direct bandgap. Ge nanoparticles exhibit Purcell factors of $2 \times 10^4$ with high radiative ($\approx95\%$) and photon ($\approx85\%$) efficiencies at the technologically relevant \SI{1.55}{\um} wavelength (Supplementary Fig.~S3). Relative to a previously proposed~\cite{rogobete2007design} infrared antenna with similar efficiency, this Purcell factor is 10 times higher.

We further demonstrate \emph{plasmon} generation~\cite{gan2012proposal} with high efficiency by using an optically thin ($t=\SI{5}{\nm}$) metal layer [Fig.~\ref{emission}(d)]. The thin metal improves the modal overlap between the gap and propagating plasmons~\cite{yang2016optically}. The Purcell factors exceed 10$^4$ for all the modes in Fig.~\ref{emission}(e). Similar to the thick-metal case, high total quantum efficiencies are achieved, with that of the $(1,1)$ mode still ${>}$90\%. Contrary to the thick-metal case, photon emission is suppressed while plasmon emission is strongly boosted: the plasmon efficiency exceeds 60\% for each of the $(1,1)$, $(1,2)$ and $(1,3)$ modes. The guided-plasmon propagation pattern [Fig.~\ref{emission}(f)] reveals highly directional plasmon launching.

\section{Widely Varying Quality Factors}
The quasistatic properties of metals~\cite{PhysRevLett.97.206806} limit the quality factors of conventional plasmonic resonances (typically ${<}$100 in the optical regime), imposing severe restrictions on many plasmonic applications. In contrast, dielectric--metal resonances provide control over the individual absorptive- and radiative-loss rates, providing options along the entire continuum between the all-metal and all-dielectric extremes.

\begin{figure}[hbtp]
  \centering
  \includegraphics[width=\linewidth]{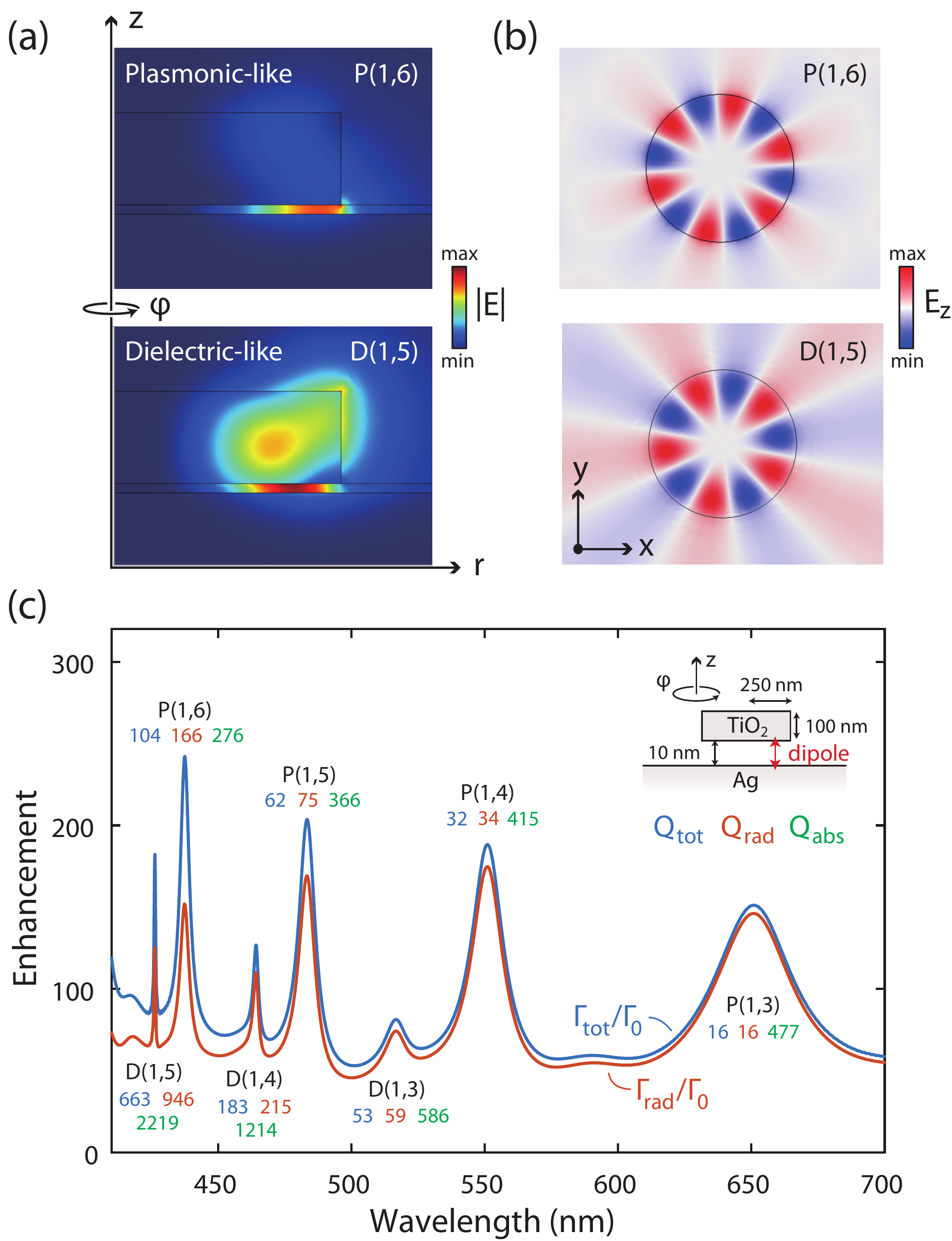}
    \caption{\textbf{Low- and high-order (whispering-gallery-like) hybrid resonances offer a large continuous design space for plasmonic quality factors.}
 (a--b)~Field profiles of the plasmonic-like [P(1,6)] and dielectric-like [D(1,5)] resonances in the (a) $r$-$z$ and (b) $x$-$y$ planes. $E_z$ are evaluated in the middle of the gap (particle) for the plasmonic-like (dielectric-like) resonance.
 (c)~Total (blue), radiative (red), and absorptive (green) quality factors of the hybrid resonances. Inset: structure and dipole excitation for quality-factor extraction.
}
  \label{highOrder}
\end{figure}

Using approximately lossless dielectrics, such as TiO$_2$ at visible frequencies, plasmonic modes with extraordinarily high quality factors can be designed (Fig.~\ref{highOrder}). As evidenced by their field patterns [Fig.~\ref{highOrder}(a--b)], the modes of the dielectric--metal resonator partition into dielectric-like and plasmonic-like resonances---both of which display strong field confinement within the gap. Figure~5(c) shows the total, radiative, and absorptive quality factors ($Q_{\text{tot}}$, $Q_{\text{rad}}$, and $Q_{\text{abs}}$) of the resonances (Supplementary~VIII). The dielectric-like modes generally have higher $Q_{\text{abs}}$ than the plasmonic-like modes because of their larger field intensity in the interior of the dielectric [Fig.~\ref{highOrder}(a)]. Unlike conventional plasmonic modes, for which $Q_{\text{tot}}$ is mainly limited by material loss, here $Q_{\text{tot}}$ is primarily limited by radiation loss, which can be readily tailored via the nanoparticle geometry and size. The $Q_{\text{tot}}$ of these resonances ranges widely from ${\sim}10$ to ${\sim}10^3$, offering a wide, continuous design space for narrow- or broad-band plasmonic applications.

\section{Robustness to Plasmonic Quantum Corrections}

Quantum phenomena beyond the classical description set the ultimate limitations on the achievable response in plasmonic nanostructures. Chief among these phenomena are nonlocality, spill-out, and surface-enabled damping~\cite{feibelman1982surface}.
In Ag, their joint impacts are well-described by a nonlocal, effective model---GNOR~\cite{mortensen2014generalized} (Supplementary~IX), a convective-diffusive hydrodynamic model---causing spectral blueshifting and broadening in structures with nanoscale features. In comparison, analogous quantum corrections in dielectrics are negligible due to the absence of free carriers. We show here that the dielectric--metal resonances display increased robustness to these detrimental quantum corrections compared to their metal--metal counterparts; taking field enhancement as a measure, the former is even superior for gaps ${\lesssim}$ \SI{5}{\nm}.
\begin{figure}[hbtp]
	\centering
	\includegraphics[width=\linewidth]{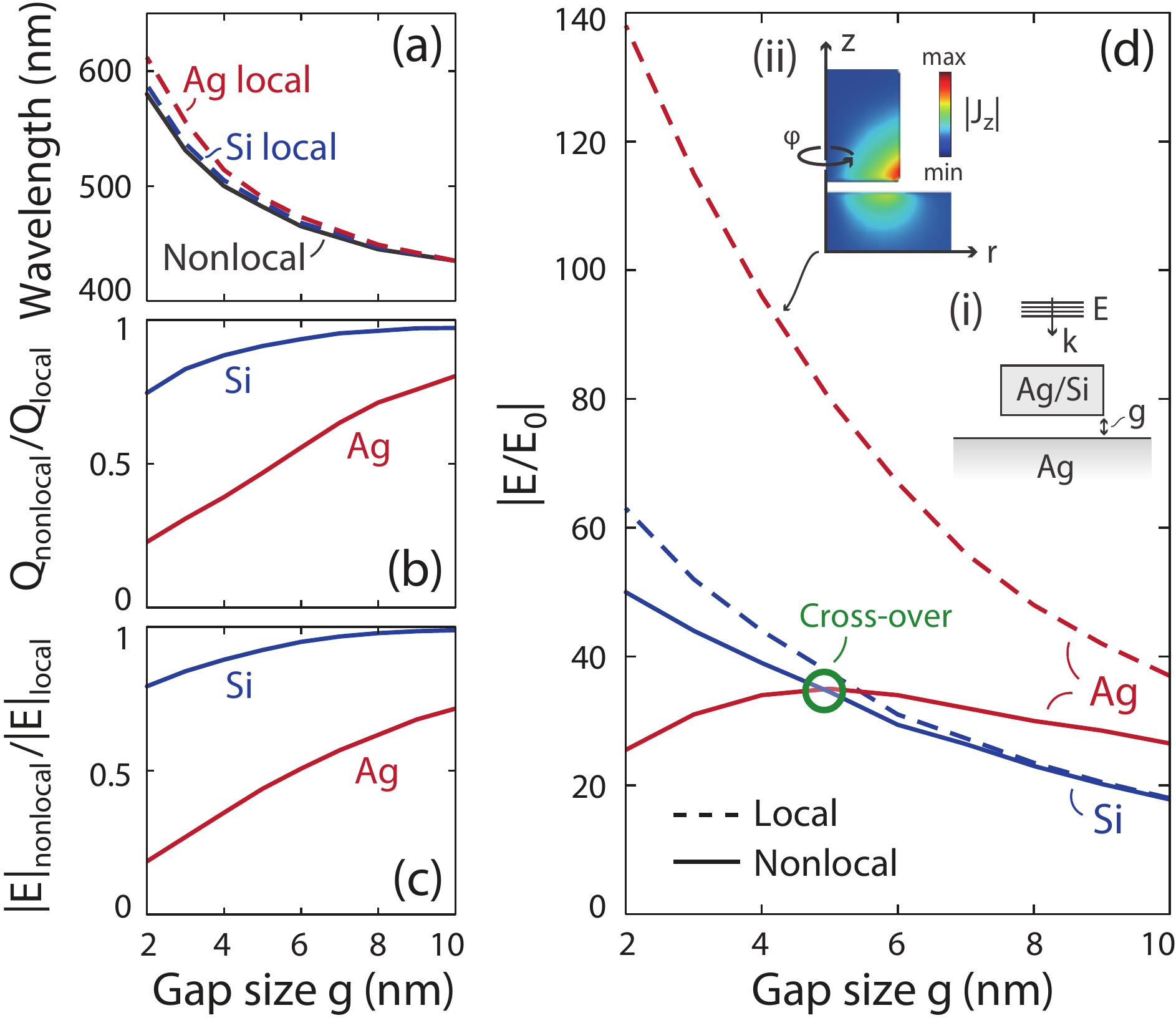}
	\caption{\textbf{
            Hybrid resonances show increased robustness to the detrimental effects of quantum corrections than their metal--metal counterparts.} The (1,1) resonances of Ag or Si nanocylinders above a semi-infinite Ag film, separated by a finite gap [inset~(i)]. The radius (height) of the Si cylinder is $50\text{ nm}$ ($40\text{ nm}$). The Ag cylinder is of identical height but of variable radius, $24\text{--}34\text{ nm}$, to spectrally align the distinct structures' (nonlocal) resonance wavelength. An effective nonlocal model~\cite{mortensen2014generalized} reveals that (a)~spectral blueshifting, (b)~linewidth broadening, and (c)~field enhancement (at gap center) reduction, relative to classical (local) predictions, are greatly mitigated in the hybrid resonators relative to metal--metal resonators.
		(d)~Accounting for nonlocal response, hybrid resonances exhibit higher field enhancement than the metal--metal resonance for gap sizes ${\lesssim}$ \SI{5}{\nm} (crossover in green marker). Inset (ii), the induced current distribution, $|J_z|$, of the metal--metal resonance (gap, $g=4\text{ nm}$).}
		\label{nonlocal}
\end{figure}

Figure~\ref{nonlocal} examines these quantum corrections for \SIrange{2}{10}{\nm} gap sizes, where inter-surface electron tunneling is absent~\cite{zhu2016quantum}. For both dielectric--metal and metal--metal structures (with equal nonlocal resonant frequencies), the resonant wavelength, quality factor, and field enhancement of the $(1,1)$ resonance are shown [Fig.~\ref{nonlocal}(a-c)] as functions of gap size. Relative to local, classical predictions, both configurations exhibit blueshifted resonant wavelengths and reductions in quality factor and field enhancement---all of which increase as the gap size decreases. Crucially, the metal--metal system suffers more severe reductions than its counterpart. This observation can be attributed to two cooperating effects: first, in light of the plasmon--Bessel framework laid out above (Fig.~\ref{analytics}), the planar multilayer equivalent approximately dictates the gap-dependent impact of quantum corrections. Accordingly, since the surface plasmon of the planar metal--dielectric--metal system suffers increased impact of quantum corrections compared to the planar dielectric--metal system (by a factor $1+\mathrm{e}^{-kg}$~\cite{christensen2016quantum} and see Supplementary Fig.~S5), the metal--metal nanoparticle's performance is similarly reduced. Second, the metal nanoparticle's edges host sharply varying current densities [Fig.~\ref{nonlocal}(d), inset (ii)] and consequently incur large nonlocal corrections in these regions.

Strikingly, the relative robustness of the hybrid resonances to quantum corrections enables them to demonstrate larger \emph{absolute} field enhancements, for equal gap sizes ${\lesssim}$ \SI{5}{\nm} [Fig.~\ref{nonlocal}(d)], than the high-intensity, pure-plasmonic metal--metal resonators. The enhancement in the latter system deteriorates drastically at these gap sizes, due to the above-noted distinguishing aspects. The comparative robustness of the hybrid resonances suggests a pathway to stronger light--matter interactions in extreme nanoscale gaps~\cite{chikkaraddy2016single}.

\section{Discussion}
In this Article, we have shown the possibility for low-loss plasmonics by coupling low-loss dielectric nanoparticles with high-confinement metallic substrates. The hybrid dielectric--metal resonances exhibit strong and efficient scattering and near-field emission enhancements, large quality factors, and nonlocal robustness beyond those of conventional plasmonic nanostructures.

By avoiding any \emph{structured} metallic components, the architecture has practical fabrication advantages. Single- or poly-crystalline metallic films exhibit much lower losses~\cite{wu2014intrinsic,mcpeak2015plasmonic} than metallic nanoparticles (which are typically amorphous, with more severe surface roughness). Moreover, this approach avoids the use of any metallic corners or tips that may strongly absorb due to fabrication imperfections.

The approach to high efficiency presented here can work in tandem with future material improvements. Just as we have shown that re-architecting common materials can improve their plasmonic response, new, low-loss materials should be integrated into these hybrid geometries rather than conventional all-metal structures. Graphene sheets behave optically very much like ultrathin metallic films, and thus our approach extends to dielectric-on-graphene architectures for efficient graphene plasmon confinement.

Looking forward, the dielectric-metal approach prompts two directions for new exploration. First, the strong emission enhancement of the dielectric--metal resonances rely on the high index contrast between the dielectric scatterer and the dielectric spacer (comprising the gap). When the index contrast is reduced, the high efficiencies can be maintained though at the expense of reduced optical confinement. Thus continued development of very-low-index ($n\approx1$) materials, such as low-index SiO$_2$~\cite{xi2007optical}, aerogels~\cite{sun2008enhanced}, and low-index polymers~\cite{groh1991lowest}, would further increase enhancements and improve efficiencies. Second, quantum effects in dielectric and dielectric--metal structures at few-nanometer length scales are of increasing interest, and should be explored further with alternative (e.g., time-dependent density functional theory) electronic and optical models. The prospect of dielectric--metal structures that are robust to deleterious nonlocal effects is especially enticing for the growing field of quantum plasmonics~\cite{fitzgerald2016quantum}.
\section*{Methods}
Methods and any associated references are available in the Supplementary Information.
\section*{Acknowledgements}
The authors thank fruitful discussions with Prof. Koppens and Dr. Bo Zhen. This work was partly supported by the Army Research Office through the Institute for Soldier Nanotechnologies under contract No. W911NF-13-D-0001. Y.Y. was partly supported by the MRSEC Program of the National Science Foundation under Grant No. DMR-1419807. O.D.M. was supported by the Air Force Office of Scientific Research under award number FA9550-17-1-0093. T.C. was supported by the Danish Council for Independent Research (Grant No. DFF¨C6108-00667). M.S. was partly supported (reading and analysis of the manuscript) by S3TEC, an Energy Frontier Research Center funded by the U.S. Department of Energy under grant no. DE-SC0001299.
\section*{Author contributions}
M.S. conceived the project. Y.Y., O.D.M., and T.C. developed the analytical models and theoretical analysis. J.D.J. and M.S. supervised the project. All authors contributed to
discussions and manuscript writing.
\section*{Additional Information}
Correspondence and requests for materials should be addressed to Y.Y. and O.D.M.
\section*{Competing financial interests}
The authors declare no competing financial interests.

\bibliographystyle{naturemag}
\bibliography{dielectricGap}


\end{document}